# Understanding The Effects of Geotechnical Properties on Viscous Erosion Rate from Plume Surface Interactions

B. Dotson, A. St. John, R. Hall, D. Sapkota, D. Britt, and P. Metzger, University of Central Florida, Department of Physics, 4000 Central Florida Boulevard, Orlando, FL 32816; email: br339171@ucf.edu.

With humans returning to the Moon under the Artemis program, understanding and mitigating effects from Plume Surface Interactions (PSI) will be essential for the protection of personnel and equipment on the Moon. To help characterize the underlying mechanics associated with viscous erosion and crater formation, experimental measurements using regolith simulants and subsonic, non-reacting flows were completed using compressed air in a splitter plate, plume cratering setup. More specifically, these investigations examined the underlying effects of bulk density, cohesion, and exhaust flow characteristics on viscous erosion rates and crater formation using Lunar highlands simulant (LHS-1), Lunar mare simulant (LMS-1), LHS-1D (Dust) simulants, and 40-80 µm glass beads in atmosphere. Results show that particle size distribution can ultimately influence crater shapes and erosion rates, likely owing to internal angle of friction. Measurements show that increasing bulk density, especially from an uncompacted to a slightly compacted state, decreases erosion rate by as much as 50%. While cohesion of granular material can mitigate erosion rates to some extent, higher levels of cohesion above 1,000 Pa may actually increase viscous erosion rates due to particle clumping. A modified version of Metzger's (2024a) equation for volumetric erosion rate is presented, with limitations discussed. These modified equations for viscous erosion, with limitations noted, show that geotechnical properties play an important role in viscous erosion and should be considered in PSI computer models for future mission planning.

1.0 INTRODUCTION

The term Plume Surface Interaction (PSI) is used to describe the complex, dynamic transport and erosion processes that happen during impingement of surface materials by rocket exhaust, particularly during launch and landing operations. During the Apollo 12 mission to the Moon, rocket exhaust from landing spacecraft ejected regolith material at high velocities, causing measurable damage to a nearby robotic Surveyor probe (Immer et al., 2011). Compared to spacecraft used during the Apollo program, the proposed size and mass of the new Human Landing System (HLS) vehicles under the Artemis program are substantially larger, creating the potential for even higher-velocity ejecta from PSI effects. Since NASA plans to develop a sustainable and long-term presence on the surface of the Moon during Artemis, complete with required lunar infrastructure, understanding and mitigating the effects of PSI events is vital for protecting personnel and equipment on the Moon.

Research into PSI events have been an on-going effort since the 1960s, with Roberts (1963), Metzger et al. (2010), as well as Rahimi et al. (2010) identifying the processes of viscous erosion, bearing capacity failure, diffused gas eruption, and diffusion driven flow as contributing to crater formation and development. Since the 1960s, other studies have leveraged a wide range of models, simulations, laboratory, and field experiments, as well as video analysis of actual lunar landing events to help study the underlying principles associated with PSI events (Fontes & Metzger, 2022; Immer et al., 2008; Immer & Metzger, 2010; Metzger et al., 2010; Metzger et al., 2008, 2011). While earlier experiments and models focused on specific destinations or space missions, more recent studies have attempted to systematically examine PSI principles, including effects of gravity and atmosphere. However, these studies appear to fall short in their systematic treatment and characterization of surface properties including particle size distributions, cohesion and bulk density on the underlying properties associated with PSI cratering events. This study examines the influence of these geotechnical properties on the development of PSI crater geometries and viscous erosion rates using small-scale experiments in atmosphere.

1.1 VISCOUS EROSION RATE

As the only significant mechanism of PSI expected on the Moon with the potential to eject high-velocity particles (Rahimi et al., 2020), determining the underlying equations associated with viscous erosion rate is critical for modeling and simulations for future lunar landings. Initially, Roberts (1963) theorized that viscous erosion of lunar regolith during the Apollo landings would be directly proportional to the shear stress of the exhaust gas, limited by the gas momentum transfer to liberated grains. While the initial equations from Roberts are still being used to predict erosion during future lunar landings, Metzger (2024a) recently showed that the theory and equations developed by Roberts may be incorrect. Leveraging analysis of Apollo landing footage in a separate study, Metzger (2024b) showed that viscous erosion on the Moon, particularly where saltation of granular particles is infrequent, is likely related to energy flux instead. Metzger (2024a) theorized that viscous erosion was proportional to the kinetic energy of the gas divided by the combined potential energy and cohesive energies of the surface, leading to the development of an improved equation for viscous erosion rate as shown in eq. 1. $\dot{V}$ is volumetric erosion rate from viscous erosion, $K$ is an unknown constant with units of velocity, $\rho$ is the gas density, $v$ is the gas jet velocity at the nozzle exit, A is the nozzle exit area, $\rho_m$ is the bulk density of the soil, $g$ is gravity, $D$ is the mean particle diameter of grains, $\alpha$ is the cohesive energy density of the soil, and $\beta$ is related to the slope of the surface ($\theta$) as shown in eq. 2.

$$\dot{V} = K \frac{\rho v^2 A}{\rho_m g D \beta + \alpha} \qquad (1)$$

$$\beta = \cos\theta \approx \frac{1}{2} \qquad (2)$$

Unlike previous measurements by Metzger et al. (2010) which used glass beads and sand without compaction, this study leverages small-scale plume experiments with high-fidelity lunar regolith simulant to examine the variables and relationships in eq. 1 in atmosphere, directly investigating any impacts from regolith density. More specifically, measurements in this study examine the influence of $D$, $K$, $\alpha$, and $\rho_m$ terms on erosion rate, as defined in eq. 1, with simulated lunar regolith. Empirical data from this investigation can ultimately improve PSI models for future lunar landings and launches. Understanding these relationships will also be helpful for mitigating viscous erosion and risks associated with PSI events.

2.0 METHODS

To determine the effect of geotechnical measurements on erosion rate and crater formation from PSI events, regolith simulants were exposed to subsonic, non-reacting flows of compressed air in a small-scale, splitter plate setup, as shown in Figure 1. This setup features a 24.8 cm x 19.7 cm x 19.1 cm aluminum regolith bin, with a clear acrylic splitter plate on one side for camera observations during PSI measurements. A 600-liter, high-pressure steel cylinder of compressed atmospheric air, with a maximum pressure of 19,305 kPa, provided uninterrupted flow through a 20,684 kPa, single-staged, pressure regulator when connected to the experimental setup. This compressed air regulator was then connected to an inline Measureman digital pressure gauge (with a reported accuracy of +/- 6.9 kPa) and purge valve, to a 500 cm flexible high-pressure hose (with an inner diameter of 2.5 cm), before passing through a Parker Skinner electronic solenoid valve (with an inner diameter of 0.32 cm), a Siargo inline mass flow gauge (with an inner diameter of 1.3 cm and reported accuracy of 0.2 slpm), and finally through a 200 cm-long, aluminum pipe (with an inner diameter of 1.0 cm). The pipe was used as the exhaust outlet, without a converging-diverging exit nozzle or device. The exit nozzle was fixed at a height of 5.0 cm above each regolith sample.

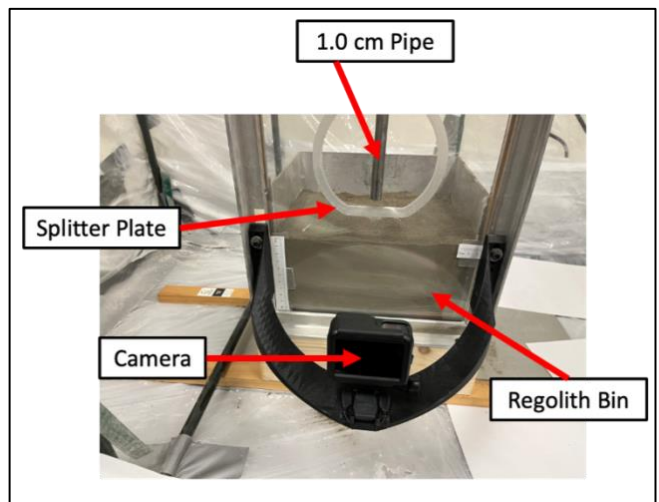

*Figure 1. Splitter plate experimental setup with regolith bin.*

Prior to measurements with regolith simulants, exhaust velocity profiles were directly measured using pitot-static devices as a function of exhaust height, and desired mass flow was accomplished by adjusting upstream chamber pressures. These velocity measurements were taken using a 10.0 cm pitot-static tube (with an inner diameter of 0.15 cm), connected to a XGMP3v3 pressure differential sensor and Arduino Uno to characterize the flow exiting the collimation pipe with a sample rate of 1,000 Hz. Velocity measurements were taken across multiple X-Y planes at various heights, as shown in Figure 2, using 3-dimensional manual translation stages to change pitot-static tube positions relative to the pipe exit in 0.1 cm increments.

From these initial measurements, when conducting PSI experiments with regolith simulants, upstream chamber pressures were set between 275 – 551 kPa using the purge valve and inline digital pressure gauge. Before each PSI measurement, the ambient atmospheric pressure was also recorded using the static pressure port on the same pitot-static device. The ambient atmospheric temperature was recorded using a Huato HE804 thermocouple probe, with a reported accuracy of 0.8 °C.

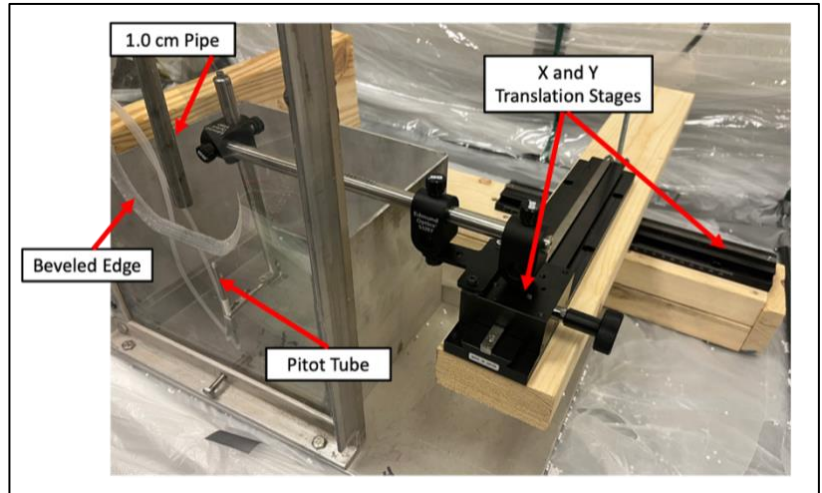

*Figure 2. Pitot-static flow characterization setup.*

For the examined plume tests, regolith simulants of LHS-1, LMS-1, LHS-1D, and 40-80 µm glass beads were individually mixed and then evenly sprinkled inside the regolith bin using a handheld scoop, taking care to avoid any inadvertent compaction or excessive clumping of material. The particle size distributions and particle shapes, as well as cohesion and shear strength as a function of normal load as reported by Dotson et al. (2024), were leveraged to provide insight into the geotechnical properties as the exact same sample materials from that previous study were examined during PSI investigations. Cohesion, *C*, was calculated for each examined sample by applying the exponential fit parameters, $A_c$ and $k_c$, for eq. 3 at measured sample bulk densities. The exponential fit parameters and average particle sizes (D), as previously reported by Dotson et al. (2024), are shown in Table 1 for LHS-1, LMS-1, LHS-1D, and 40-80 µm glass beads. Additional information regarding these high-fidelity lunar simulants can be found at: https://spaceresourcetech.com.

$$C = A_c\ e^{k_c \rho_m} \qquad (3)$$

*Table 1. Exponential fit parameters for calculating cohesion (in Pa) as a function of density.*

| Sample | $A_c$ | $k_c$ | D (µm) |
|---|---|---|---|
| LHS-1 | 0.45 ± 0.36 | 4.46 ± 0.40 | 92 |
| LHS-1D | 2.12 ± 1.34 | 5.35 ± 0.33 | 30 |
| LMS-1 | 0.27 ± 0.09 | 4.42 ± 0.31 | 90 |
| 40-80 µm glass beads | 0.39 ± 0.17 | 4.35 ± 0.33 | 53 |

To initially estimate bulk density, the mass of material placed inside the regolith bin was recorded using a Fristaden digital scale (with published accuracy of 0.01 g), while the volume of the examined regolith simulant was recorded using 3-dimensional LiDAR scans of the bin, before and after filling with the ScannerPro app on an iPad Pro (6th generation). A horizontal "fill line" was placed along the inside of the regolith bin using white tape, in order to maintain a consistent volume and separation from the pipe exit. For PSI measurements at higher bulk densities, the regolith simulant was evenly sprinkled above the "fill line" and bulk density was then increased using a 12V electronic vibration motor affixed to the side of the aluminum regolith bin, opposite the clear acrylic splitter plate. With the entire regolith bin shaking, mechanical vibrations at rates between 20-40 Hz were used for various durations between 5 – 300 seconds, in order to compact the regolith simulants so that the surface was even with the horizontal "fill line", resulting in higher bulk densities.

The density of each regolith simulant was directly measured inside the regolith bin, prior to PSI flow measurements, using gamma-ray spectroscopy. Adapting the techniques developed by Santo & Tsuji (1977), a custom, removable, gamma-ray spectroscopy setup was used to non-destructively measure the attenuation of gamma-rays through the regolith bin and regolith sample prior to each run. This setup used a 10 µCi sample of radioactive Cesium-137 to emit gamma-rays that could be measured using a 3.8 cm x 3.8 cm NaI(Tl) crystal gamma ray scintillation detector (with a detectable range up to 10,000 keV), separated by a vertical translation stage (with 0.1 cm markings). Before taking sample measurements, background gamma ray spectra were taken at a fixed separation distance (without the sample or regolith bin), as well as with the empty aluminum regolith bin and regolith bin with regolith simulant inside. Gamma-ray spectra were recorded using an observation time of 600 seconds and bin width of 0.0001 keV, before being analyzed using the Pulse Recorder and Analyzer (PRA) software, and curve fitting with a 3-5 term gaussian applied to the characteristic monatomic spectral feature near 662 keV for radioactive Cesium-137 in Matlab. This removable, gamma-ray, density measuring setup was calibrated by measuring samples of LHS-1, LMS-1, LHS-1D, and 40-80 µm glass beads at known densities (from mass and volume) in both a small-scale regolith container, as well as an aluminum regolith bin of identical scale and design. For all measurements, a minimum of 3 separate spectra, with observation times of 600s, were taken for each density and averaged.

After sample preparation and density measurements, non-reacting flow of compressed air was initiated using a 12V switch, connected to the electronic solenoid valve, with corresponding pressure and mass flow measurements measured in-line. The exhaust flow was then split by the clear acrylic splitter plate so the side profile of PSI crater development could be recorded using a GoPro Hero 7 camera with a frame rate set to 60 and 240 fps. The splitter plate has a 2.5 cm beveled edge, which angles away from the regolith bin at a 45 deg angle, and a straight edge on opposite side as shown in Figure 2, to reduce any disturbances to the impinged surface flow as much as possible. Monitoring crater formation with various regolith simulants at a set bulk density, video tracking of the crater shape and ejected material was completed using image analysis techniques developed in Python to determine the crater depth (H), crater width (w), and crater volume (V) as a function of time throughout the PSI event. This image analysis technique leveraged edge detection for extracted frames to trace the outline of the crater shape and depth over time, as shown in Figure 3. Similar crater depth tracing was performed manually using Engauge Digitizer software for each run, examining frames every 0.5 seconds, to confirm the correct output of the automated Python edge detection analysis. By placing scale bars and a grid pattern on the front of the regolith box, crater depth was determined using this manual tracking and then compared to automated edge detection outputs (Figure 3) to verify correct output. The automated edge detection software assumes a 3-dimension rotation of the 2-dimension image to determine volume; thus, assuming a rotationally symmetric crater shape, consistent with previous data processing from Metzger (2010). Using this technique, crater depth and volume were plotted as a function of time using 5-σ data clipping in Python to remove any outlying points from airborne dust, which caused small time history gaps in Figure 3b. The volumetric erosion rate was then calculated by linear fitting of these clipped time histories for each run under steady-state conditions, between 4 and 6 seconds into the run as shown in Figure 3b.

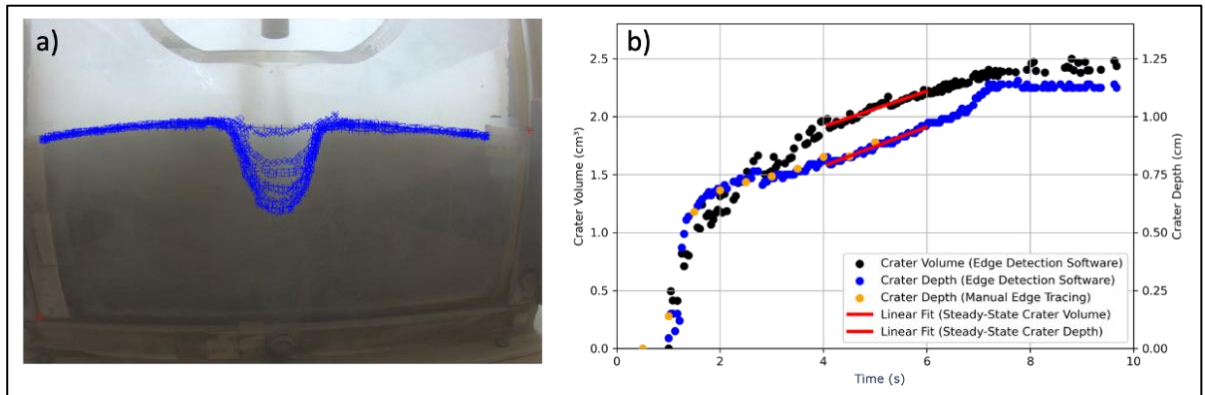

*Figure 3. a) Image edge detection of crater geometries over time b) Output of crater depth and volume over time.*

3.0 RESULTS

3.1 Plume Characterization

Pitot tube measurements of normal (downward) velocity, recorded in different planes below the nozzle exit inside the empty regolith bin are shown in Figure 4 for an upstream chamber pressure of 551 kPa and mass flow rate of 129 standard liters per minute (slpm). In Figure 4, the exit nozzle of the 1.0 cm pipe is centered at 0 cm in both the X and Y positions, as shown. These measurements reveal that the gas velocity maintains a sub-sonic, collimated profile between 5,000-5,500 cm/s directly under the exit nozzle to a depth of roughly 2.0 cm, before slowly expanding radially outwards. Moreover, a downward, normal velocity on the order of 5,000 cm/s was observed at 7.5 cm below the exit.

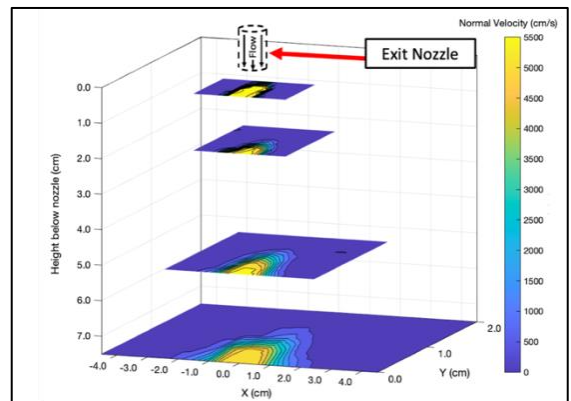

*Figure 4. Flow velocity measurements below exhaust nozzle (551 kPa, 129 slpm).*

From this velocity profile of the gas, there does not appear to be any notable interactions or reduction of flow from the splitter plate (located along the 0 cm Y-position in Figure 4). While the velocity profile will change as a regolith surface and crater are introduced into the flow, these initial measurements can also be used to calculate the Reynolds number ($Re$) associated with the setup using eq. 4 below, where $D_p$ is the pipe diameter, $V$ is the flow velocity, and $\mu$ is the dynamic viscosity (Sommerfeld, 1908).

$$Re = \frac{\rho v D_p}{\mu} \quad (4)$$

Assuming a 1.0 cm diameter pipe, as well as a density of 0.001225 g/cm³ and dynamic viscosity of roughly 0.000018 Pa·s for room temperature air, it is clear from velocity measurements that the non-dimensional Reynolds number for this flow is roughly Re ≈ 37,417; consistent with turbulent flow (Re > 2,900). Similar measurements were taken for upstream chamber pressures of 275 – 551 kPa, with Reynolds numbers above 2,900 as well, also indicating a potential for turbulent flow in all examined test regions.

3.2 Crater Formation and Shapes

Analysis of plume videos reveals that only surface scouring and saltation was observed during crater formation, consistent with viscous erosion as the sole erosion process in this study. These observations match previous measurements of viscous erosion by Metzger et al. (2010), which used the same experimental setup and flow rates. Given that pitot-static measurements confirm subsonic gas flows (Section 3.1), it is unlikely that bearing capacity failure is present during trials, especially as surface normal forces and velocities are small. Additionally, no other particle transport phenomenon were observed during plume measurements, such as tangential granular flows were observed which would otherwise suggest bearing capacity or diffusion driven flow. Since crater depth and volume measurements are taken throughout plume impingement, diffused gas eruption was not observed in this study as well, especially since this erosion mechanism only occurs after removal of high-velocity impinged flow (Metzger et al., 2010). Taken together, these observations suggest that viscous erosion is the only plausible mechanism for all examined measurements.

As shown in Figure 3b, crater depth and volume time traces had similar steady-state slopes, but depth always reached steady-state first for each sample examined. In addition to measuring the depth and volume changes for each run, differences in overall crater shapes were noted for impingements with regolith simulant and the 40-80 µm glass beads. As shown in Figure 5a below, a compound crater shape with two different radii was produced using a mass flow rate of 215 slpm impinged on 40-80 µm glass beads, similar to results previously reported by Metzger (2024a) for other glass bead sizes. However, using the same setup and mass flow rate, flow impinged on uncompacted LMS-1 produced a more-simple crater shape with steep walls (Figure 5b). The singular crater shape shown in Figure 5b was also observed with compacted LMS-1, and for runs with LHS-1 and LHS-1D at all levels of compaction. The compound crater shape in Figure 5a was only observed with glass bead trials and the potential mechanism behind these shape differences is discussed in Section 4.2.

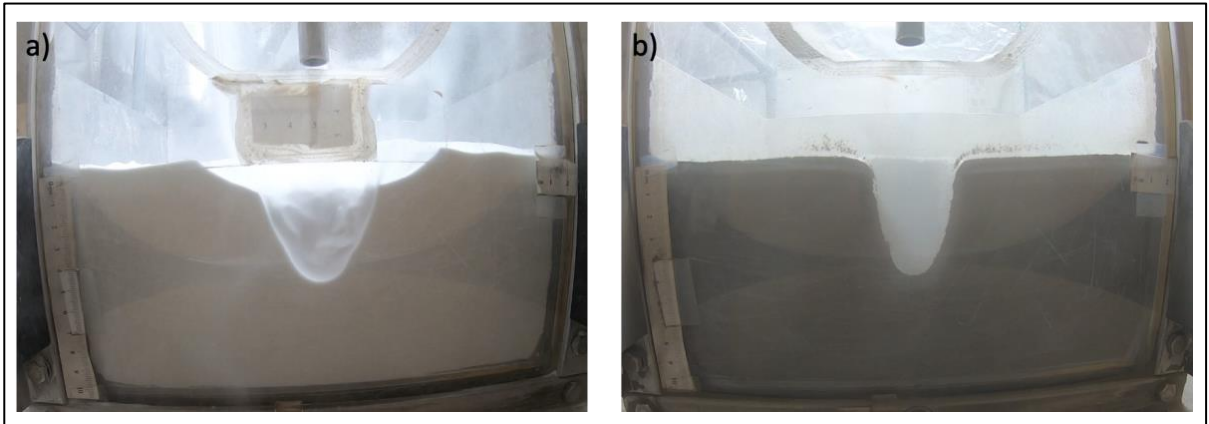

*Figure 5. Crater cross section of a) 40-80 μm glass beads and b) LMS-1 without compaction; Both craters were formed with similar ambient conditions and a constant mass flow rate of 215 slpm.*

3.3 Volumetric Erosion Rate and Bulk Density

From gamma ray attenuation measurements, the observed ranges of bulk density during this study were 0.87-1.16 g/cm$^3$ for LHS-1D, 1.34-1.86 g/cm$^3$ for LHS-1, 1.60-2.03 g/cm$^3$ for LMS-1, and 1.56-1.62 g/cm$^3$ for 40-80 μm glass beads. The volumetric erosion rate, calculated from linear fitting of time histories, is shown in Figure 6 as a function of bulk density and mass flow rate for LHS-1, LMS-1, LHS-1D, and 40-80 μm glass beads. The vertical error bars represent the 1-σ error calculated from repeat trials, while horizontal error bars have been omitted for clarity. Density has been measured with a resolution of 0.07 g/cm$^3$ for each run.

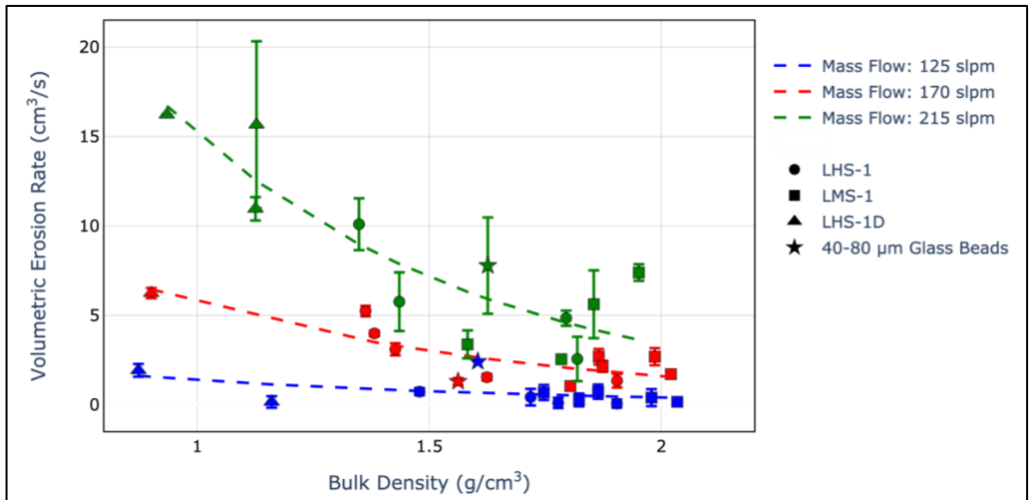

*Figure 6. Volumetric erosion rates as a function of bulk density and mass flow rate.*

The average observed volumetric erosion rates ranged between 2.5-16.2 cm$^3$/s for trials with a constant mass flow rate of 215 slpm, 1.0-6.2 cm$^3$/s for trials with a constant mass flow rate 170 slpm, and 0.1-2.4 cm$^3$/s for trials with a constant mass flow rate of 125 slpm. In Figure 6, dashed lines have been added between data points to highlight the general pattern in the measurements with respect to mass flow rate. These lines are intended for illustrative purposes and do not represent any specific model or mathematical relationship between the variables. While volumetric erosion rate generally decreases with an increase in bulk density for all mass flow rates, the influence of bulk density on volumetric erosion rate was most pronounced for increasing mass flow rates. There does not appear to be any trends based on regolith type or particle size in Figure 6, with mass flow rate and density noted as the dominate parameters.

3.4 Volumetric Erosion Rate and Cohesion

    Leveraging the equation for cohesion as a function of density (eq. 3), as well as the empirical fit values in Table 1 for LHS-1, LMS-1, LHS-1D, and glass beads from Dotson et al. (2024), the volumetric erosion rate is shown in Figure 7 as a function of cohesion. Of note, both axes in Figure 7 have been plotted using a logarithmic scale in order to highlight relationships across a broad range of cohesion values and volumetric erosion rates from viscous erosion. Calculated cohesion values ranged between 184-3,062 Pa for trials with a constant mass flow rate of 215 slpm, 196-2,202 Pa for trials with a constant mass flow rate of 170 slpm, and 324 – 4,124 Pa for trials with a constant mass flow rate of 125 slpm. As shown in Figure 7, dashed lines are drawn between points of similar mass flow rate for illustration purposes only and do not represent a specific model or mathematical fit. For all mass flow rates, increasing simulant cohesion initially resulted in a decrease in observed volumetric erosion rate. Yet, continuing to increase cohesion eventually resulted in no change or a slight increase in volumetric erosion rate. The largest change in observed volumetric erosion rates was noted for mass flow rates of 125 slpm. This behavior is discussed in Section 4.4 below. There does not appear to be any trends based on regolith type or particle size in Figure 7, as data points are well distributed.

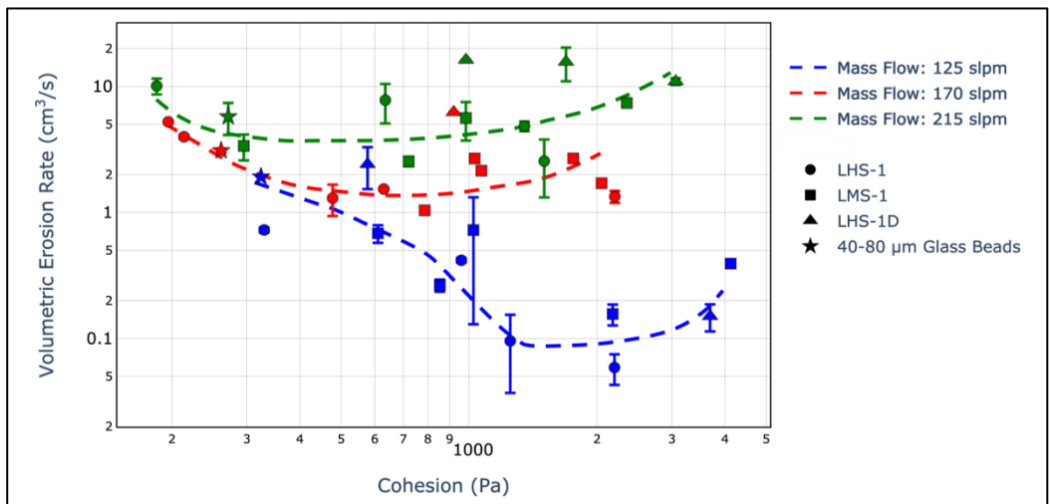

*Figure 7. Observed volumetric erosion rates plotted against cohesion calculated using eq. 3.*

3.5 Equations for Viscous Erosion

    The observed volumetric erosion rates due to viscous erosion, as measured during this study, are plotted in Figure 8 against the calculated values for volumetric erosion rate (using eq. 1, which assumes viscous erosion) for each mass flow rate. The calculated volumetric erosion rates from eq. 1 (X-axis) ranged from 0.2-10.3 cm$^3$/s, while the observed volumetric erosion rates ranged from roughly 0.1 – 16.2 cm$^3$/s. The dashed line in Figure 8 is a linear fit with a slope of 1 and intercept of 0, representing where the X-axis and Y-axis values are equal. If the experimental results match the model in eq. 1, the data points should fall on the dashed line as shown. In Figure 8, the observed volumetric erosion rates from viscous erosion generally match the theory well, with a goodness of fit ($R^2$ value) of 0.72, with some discrepancies discussed in

Section 4.5. There does not appear to be any trends based on regolith type or particle size in Figure 8 either, as data points are well distributed around the dashed trendline.

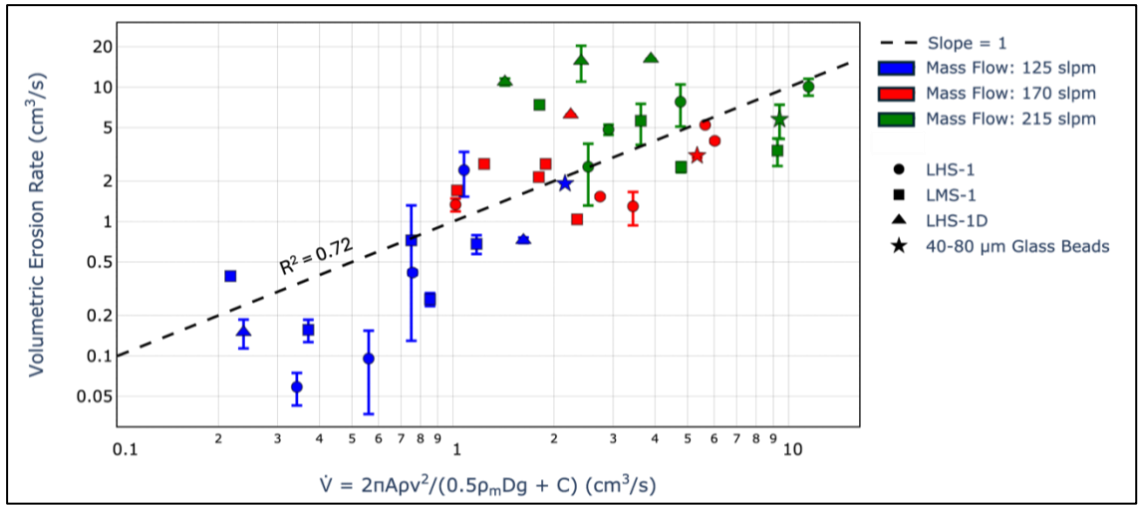

*Figure 8. Observed volumetric erosion rates plotted against the calculated volumetric erosion rate due to viscous erosion (eq. 1) with constant K = 2π, β = 0.5 and α = Cohesion, C; Dashed line represents where observed rates and calculated rates from eq. 1 are equal (line with a slope of 1).*

Of note, while $\alpha$ in eq. 1 has been theorized to represent cohesive energy density (in units of J/m$^3$), the X-axis values in Figure 8 have been calculated using cohesion, $C$, from eq. 3 (in units of Pa), which ultimately reduces to the same units. Additionally, when calculating rates for volumetric erosion (X-axis) in Figure 8, the constants $K$ and $\beta$ have been set to 2π cm/s and 0.5, respectively. This assumed value for β=0.5 is taken from previous calculations by Metzger (2024a), while $K$ was determined calculating a goodness-of-fit ($R^2$) between the observed data to the dashed line in Figure 8 (slope of 1, intercept) for various values of $K$. As shown in Figure 9, the best good-of-fit ($R^2$) value was calculated to be 0.72, corresponding to $K \approx$ 6.3 cm/s $\approx$ 2π cm/s.

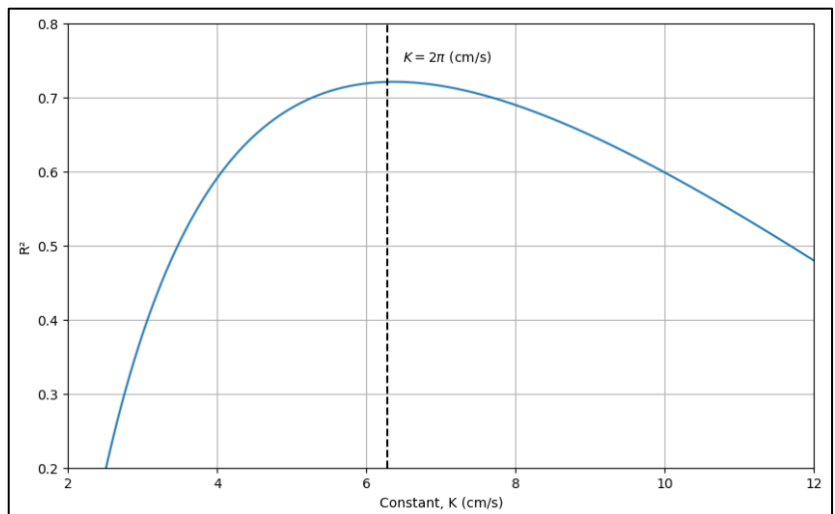

*Figure 9. Relationship between goodness-of-fit ($R^2$) and values for constant, K (eq. 1) when fitting observed volumetric erosion rates; As shown in Figure 8, the best fit of observed volumetric erosion rate data occurs when K = 2π cm/s, which corresponds to $R^2$ = 0.72.*

## 4.0 DISCUSSION

### 4.1 Flow measurements

As discussed in Section 3.1 above, the calculated Reynolds number for the measured exhaust velocities is usually associated with a fully developed turbulent jet. Such a turbulent jet will have irregular and chaotic mixing that could lead to larger boundary layers when impinging on a surface or when flowing along the splitter plate. However, previous observations by Phares et al. (2000) suggests that impinged flows can remain laminar, even with high Reynolds numbers, until several times the exit pipe diameter. Thus, additional measurements of flow characteristics across the impinged surface profile are needed to determine where turbulence may be triggered in this setup, or if the flow remains laminar in the local regime. Nevertheless, the measurements in Figure 4 do not reveal any observable surface boundary effects from the splitter plate wall, at least given the spatial resolution of measurements taken. Such a boundary interaction must therefore be less than 0.1 cm, demonstrating that the splitter plate design is adequate for experiments with similar exhaust velocities and geometries. Future studies may further examine the effects of splitter plate shape and placement on the resultant flow characteristics.

Based on pitot-static measurements, small downward velocities of 100-200 cm/s were observed at the exit plane, at a radial distance of roughly 0.1-0.3 cm. This observed flow likely suggests measurement of a small, entrained flow, in the same direction as the exhaust jet. While this entrained flow is not likely to contribute significantly to the normal force on an impinged surface, future studies should consider flow entrainment when dealing with higher exhaust velocities. While flow entrainment will be insignificant on Mars, or non-existent on the Moon, large downward velocities from flow entrainment may alter the flow characteristics and velocity profiles when landing large rockets on planetary bodies with an atmosphere.

The exhaust velocity measurements directly below the nozzle show a well-collimated jet of gas that expands radially as the plume travels farther from the exit plane. For the examined upstream chamber pressures and nozzle geometry, the centerline exhaust velocity slows by approximately 5-10% at a distance of roughly 7.5 times the nozzle diameter from the exit plane, namely when exposed to atmosphere and without an impinged surface. The exhaust velocity profile expands radially to roughly 4 times the nozzle radius at a similar distance from the exit plane, inducing radial velocity components as the plume expands. When exposed to an impinged surface, this well-focused flow exhibits a high normal stagnation pressure directly on the exhaust centerline that decreases radially, consistent with previous measurements by Fontes et al. (2022). Taken together, these observations of a focused and well-collimated jet will likely change when exposed to reduced ambient pressures or vacuum, such as on the Moon or Mars. For reduced ambient pressures, the flow is expected to expand even more in the radial direction, producing a different crater shape than observed in this study. While future studies should examine the equations for viscous erosion in vacuum or reduced ambient pressures, this study provides valuable insight into the initial governing equations associated with viscous erosion and geotechnical properties of a granular surface in terrestrial atmosphere.

4.2. Crater Shapes and Internal Angle of Friction

Unlike previous plume measurements, crater geometries observed with high-fidelity regolith simulants during this study maintained a parabolic cross-sectional shape (Figure 5b), without a clearly defined inner and outer crater as observed with glass beads (Figure 5a) and as reported by Metzger (2024a). Since these previous experiments were conducted with uniform particle size distributions, it is likely that the complex distribution of particle sizes in examined lunar regolith simulants can affect the overall crater geometry and resultant erosion rates due to internal angle of friction of the granular material. Internal angle of friction is a material property that represents the angle at which granular material begins to shear or fail. As previously noted by Metzger (2024a), the primary mechanism which drives the development of a more complex, inner and outer crater shape (Figure 5a) during plume impingement is wall collapse, where the angle of the crater shape exceeds the internal angle of friction for the material. While radial flow from the plume may provide a stabilizing force on the crater walls, collapsed wall material can be recirculated by the plume and eroded via entrainment, increasing erosion rates.

Previous measurements by Dotson et al. (2024) have shown that materials with monoatomic particle size distributions, such as glass beads, generally have lower internal angles of friction which would make wall collapse during crater formation more likely. Thus, polydisperse particle size distributions, such as with the lunar regolith simulants in this study, can have higher internal angles of friction allowing for steep craters and different erosion mechanisms. Since the internal angle of friction is expected to be higher for actual lunar regolith, such a relationship implies that viscous erosion from rocket exhaust may be less on the Moon, but future PSI studies should examine other high-fidelity regolith simulants, including those with agglutinates (Li et al., 2023). Since previous measurements have demonstrated that internal angle of friction for granular materials increases with density, viscous erosion rate while landing and launching from the planetary surfaces may be somewhat reduced through compaction (Dotson et al., 2024). Measurements are already underway to calculate the observed angle of internal friction from crater slopes and geometries over time, but local variations appear to be highly dynamic, likely a result of turbulent flow discussed in the previous section. Future studies with slower, laminar flow may be better suited for such direct measurements of slope angles and the impact from internal angle of friction on viscous erosion rate.

4.3 Bulk Density

The results of this study show that viscous erosion rate decreases with an increase in bulk density of the surface material. Consistent with the relationship in eq. 1, the viscous erosion rate is inversely proportional to the bulk density of the granular material and dependent on mass flow rate, suggesting the theory from Metzger (2024a) are correct. Higher densities likely exhibit stronger interparticle forces that resist erosion, helping to reduce viscous erosion rate. For uncompacted lunar regolith simulants under atmospheric conditions, increasing the mass flow rate of impinged gas results in a higher volumetric erosion rate, as expected. For a constant mass flow rate, increasing the bulk density results in a decrease in volumetric erosion rate from viscous erosion. However, the influence of bulk density on volumetric erosion rate is reduced with a decrease in mass flow rate as lower mass flow rates generally result in smaller volumetric erosion rates overall. This can be seen in Figure 6, where increasing the bulk density for uncompacted regolith simulants by roughly 50% results in a ~50% reduction of observed volumetric erosion rates for all mass flow rates considered. However, while increasing the bulk density of the granular surface material generally reduces viscous erosion rate, the overall effect from viscous erosion is most prominent for initially uncompacted surfaces, with less of an impact at higher bulk densities. Nevertheless, when building landing or launch pads on granular material such as lunar or Martian regolith, the observed behavior suggests that surface compaction will be beneficial to help mitigate viscous erosion.

It is important to note the limitations of measurements, particularly as this study assumes a homogeneous level of compaction throughout the regolith bin, even as a function of depth. This is especially true when taking bulk density measurements via gamma ray attenuation vertically through the sample, as well as when estimating bulk density from known mass and volume. Assuming homogeneity is likely suitable for determining first order relationships with respect to bulk density and erosion rate, given the sample is only 10.0 cm deep. Nevertheless, density undoubtedly changes as a function of depth, as layers apply a compacting normal load on subsequent deeper layers. Studies are already underway that leverage cone penetrometers and collimated gamma ray beams to better understand the relationship between sample density and depth.

As shown in Figure 6, there are large error bars associated with runs with mass flow rates of 215 slpm, as this material often clumped together and became liberated explosively, lofting large chunks sporadically, before eventually achieving a steady-state erosion rate. This effect was more common with LHS-1D, owing to large uncertainties in the observed erosion rate calculations for bulk densities less than 1.2 g/cm$^3$. While the general relationship between volumetric erosion rate and bulk density appears to be the same for all observed mass flow rates, this behavior may suggest a slightly different erosion behavior for fine-grained particles with higher levels of compaction, where electrostatic charging may affect particle behavior, especially under lunar gravity conditions. Future studies should examine different particle size distributions, bulk densities, and electrostatic charging effects to better understand their impact on erosion rate and effective cohesion.

4.4 Cohesion

As shown in Figure 7, for samples with cohesion less than 1,000 Pa, additional cohesion results in a decrease in the observed volumetric erosion rates of granular material from viscous erosion which is generally consistent with the theory presented by Metzger (2024a). While Metzger (2024a) introduces the variable $\alpha$ as the cohesive energy density of a soil, the results of this study suggest that sample cohesion provides a good approximation for $\alpha$, particularly when cohesion is less than 1,000 Pa. More specifically, the results in Figure 7 demonstrate that volumetric erosion rate from viscous erosion is inversely proportional to cohesion, $C$, as shown in eq. 5 below, where a and b are constants.

$$\dot{V} \propto \frac{1}{a+b\cdot C} \qquad (5)$$

With a closer examination of Figure 7, an increase in cohesion from 200 Pa to 1,000 Pa results in a roughly 63% decrease in volumetric erosion rate for mass flow rates of 170 and 215 slpm. Similarly, increasing cohesion from roughly 300 Pa to 1000 Pa results in a roughly 86% decrease in observed volumetric erosion rate for a mass flow rate of 125 slpm. Taken together, these values can be used with eq. 5 to determine the relative size of constants *a* and *b*. For all examined mass flow rates, such a calculation reveals that *a* is roughly 1.6 times larger than *b* in eq. 5, specifically in the range where cohesion is less than 1,000 Pa. By comparing eqs. 1 and 5 directly, this suggests that the potential energy terms ($\rho_m g D \beta$) in the denominator of eq. 1 is roughly 1.6 times greater than the cohesive terms ($\alpha$, in this case set equal to cohesion).

However, for cohesion values greater than 1,000 Pa in Figure 7, a further increase in sample cohesion results in higher observed volumetric erosion rates, not predicted by previous theory. For examined mass flow rates greater than 170 slpm, increasing the cohesion from 1,000 Pa to 2,000 Pa results in roughly a 60% increase in volumetric erosion rate. For a mass flow rate of 125 slpm, an increase in cohesion from 2,000 Pa to 3,000 Pa results in a similar 60% increase in volumetric erosion rate. Using these additional values in eq. 5 to calculate relative sizes of terms reveals that *a* (similar to potential energy terms in eq. 1) is roughly 3-3.6 times greater than *b* (similar to cohesion terms in eq. 1) for cohesion values above 1,000 Pa. When comparing such relative sizes of terms, it is clear that the relationship between volumetric erosion rate and cohesion is not linear and likely depends on mass flow rate.

As flow is impinged on a granular surface, higher levels of cohesion can help to reduce the volumetric erosion rate to an extent, but its influence is ultimately reduced for samples with very high cohesion (generally greater than 1,000-2,000 Pa). This likely results from the effect of interparticle forces, which can work to bind the granular material together to resist viscous erosion or shearing forces (Lee, 1995). At higher levels of cohesion, however, tightly bound groupings of particles may be liberated from the surface in clumps, owing to an increase in the overall amount of eroded material. Yet, for lower mass flow rates (near 125 slpm), the potential energy of lifting a particle over a neighboring particle tends to dominate. Previous investigation of the lunar surface shows that inherent cohesion of the regolith can vary greatly depending on location but is generally between 200 to 1,000 Pa (Ming, 1992). Thus, the inherent cohesion lunar regolith may help to reduce viscous erosion rates when landing or launching rockets from the surface of the Moon by roughly 63-86%, for mass flow rates between 125 and 215 slpm.

While the cohesion values presented in this study were determined by leveraging the generalized equations developed by Dotson et al. (2024), these models were derived from maximum shearing force only. As such, this study did not account for dilation of granular material from shearing forces, non-linear effects of reduced gravity, or reduced normal loads which may change the overall cohesive nature of a sample. Such effects would likely cause the data in Figure 7 to change only the calculated cohesion values (X-axis) shrinking the range of observed cohesion values. Therefore, when accounting for dilation or non-linear effects, the calculated cohesion value may be less than reported in this investigation, but the overall inverse relationship to viscous erosion is expected to remain the same.

Future studies should examine the relationship between erosion rate, cohesion, and cohesive energy density as initially theorized by Metzger (2024a). To calculate cohesive energy density, a more substantial understanding of sample density and cohesion as a function of depth is required. As noted in Section 4.3 above, measurements are already underway to help characterize the change in geotechnical properties as a function of depth within a granular sample. Since density is expected to increase with sample depth, it is unlikely that the cohesive energy density profile is homogenous throughout the sample as well. Nevertheless, using density and cohesion as an initial estimate provides valuable insight into the relationship between viscous erosion rate and sample cohesion, as well as general weighting of terms in eqs. 1 and 5.

4.5 Equations for Viscous Erosion

The observed volumetric erosion rates presented in Figure 8 match the theoretical predictions from Metzger (2024a) when adjusted slightly as shown in eq. 6, where *f(C)* is an unknown non-linear function of surface material cohesion. By setting β=0.5, a reasonable first-order approximation of the erosion rate can be achieved for examined materials with cohesion values below 1,000 Pa, as shown in eq. 7 and Figure 8. Scatter in the observed data is noted in Figure 8, as observed data points don't fall exactly on the predicted line, likely due to the inherent variability of granular materials, the complexity of erosion processes, or the presence of other potential PSI mechanisms. However, the largest deviations from the model in Figure 8 are evident for cohesion values greater than 1,000 Pa, corresponding to the nonlinear influence of cohesion discussed in the previous section. While the ambient conditions during these tests differ greatly from those on the Moon, the relationship in eq. 7 is expected to still provide a basis for initial prediction of erosion rates, especially given the fact that only viscous erosion processes were observed during these trials (Section 3.2). However, additional experiments are already underway to examine the impact of ambient pressure on eq. 7 below. Since these governing equations for viscous erosion are being used in computer simulations for future Moon landings and launches, special care therefore should be given to consider these potential limitations of eq. 1 (Rahimi et al., 2020). Future research is necessary to develop a more accurate relationship between α and material properties like cohesion and density, especially as a function of depth.

$$\dot{V} = 2\pi \frac{\rho v^2 A}{\rho_m g \beta D + f(C)} \qquad (6)$$

$$\dot{V} \approx 2\pi \frac{\rho v^2 A}{0.5 \rho_m g D + C} \qquad \text{[for C < 1,000 Pa]} \qquad (7)$$

The fact that observed data seems to match theorized results best when *K* is equal to 2π may be related to the circular geometry of either the nozzle or crater shape. However, if the value 2π is purely related to geometries, then an additional variable is likely missing from eqs. 1, 6, and 7 that is roughly equal to 1 cm/s at room temperature and pressures since constant *K* is expected to have units of velocity. Metzger (2024a) theorized that the thermal velocity of gas in the boundary layer, as opposed to gas jet velocity ($v$), may be a better variable for predicting erosion rate, which may also yield a different value for constant *K*. Nevertheless, the influence of temperature variations on *K* remains uncertain, as all experiments were conducted at room temperature. Since exhaust velocities and temperatures for rockets landing or launching on the Moon or Mars are expected to be significantly higher than those observed in this study, future investigations should also consider experiments with commensurate exhaust velocities, temperatures, and surface properties.

4.6 Shear Strength

While this study did not make any attempt to directly measure the relationship between viscous erosion rate and shear strength, previous studies have shown the important relationships between shear strength, cohesion, and bulk density more directly (Dotson et al., 2023, 2024). These previous studies represented shear strength as a 3-dimensional surface that increases with an increase in density and normal load, where cohesion represents one of the edges of this 3-dimensional surface. Since viscous erosion rate generally decreases with an increase in cohesion and bulk density as described above, viscous erosion rates will likely be reduced where granular shear strength is high; generally consistent with both Roberts (1963) and Metzger (2024a). However, additional research is needed to understand the influence of these geotechnical properties, shear strength, and erosion rate, particularly for granular materials with cohesion greater than 1,000 Pa. Additionally, since Dotson et al. (2024) demonstrated that shear strength can change with absorbed atmospheric water increasing interparticle cohesion, additional plume tests should be conducted in vacuum or under dry conditions for comparison.

However, measurement of normal and shear velocities along the crater wall are required to fully understand and predict the relationship between shear strength and viscous erosion rates. This can be challenging as the crater geometry is ever-changing with time and sample density changes as a function of depth. Studies are already underway to directly measure the shear and normal force from impinging flow within the crater, leveraging 3-D prints of LiDAR scans from this study. Such investigations are expected to reveal areas of maximum erosion rate within a steady-state crater, ultimately providing insight into the flow behavior and fundamental characteristics associated with crater formation and development.

5.0 CONCLUSION

Geotechnical properties play an important role in PSI events and determining erosion rates from plumes impinged on granular material. The results of this study show that more complex particle size distributions, likely owing to higher internal angles of friction, can change the overall crater shape and may influence erosion rates. Measurements in this study also show that viscous erosion rate decreases with an increase in bulk density, particularly when transitioning from an uncompacted to a slightly compacted state. Higher levels of cohesion may help to reduce erosion rates form PSI events to an extent but can also increase erosion rates for values greater than roughly 1,000 Pa, potentially due to particle clumping. The potential energy of lifting a particle over a neighboring particle tends to dominate erosion rate over cohesion contributions, particularly for lower mass flow rates near 125 slpm. The equations for viscous erosion rate presented by Metzger (2024a) generally match experimental data from this study with a few adjustments, including using sample cohesion as an input instead of cohesive energy density. A modified version of such equations, determined from experimental data, is presented but should be limited to models where cohesion is less than 1,000 Pa. The viscous erosion rate likely decreases with shear strength, but additional tests are needed to measure normal and shear forces directly, potentially leveraging 3-D LiDAR scans and prints. Additional studies should examine the effects of gas velocities and temperatures on the rate of viscous erosion from PSI events. Future studies should also consider the change in density as a function of depth, which likely affects cohesion, shear strength, and viscous erosion rate. Understanding the limitations in the equations for viscous erosion rate are important when establishing PSI computer codes, especially when modeling upcoming missions in the Artemis program. Failure to accurately predict the influences of geotechnical properties on viscous erosion rate from PSI events may result in significant damage to equipment, vehicles, or crew.

6.0 DECLARATION OF COMPETING INTEREST

The authors declare that they have no known competing financial interests or personal relationships that could have appeared to influence the work reported in this paper.

7.0 ACKNOWLEDGMENTS

This work is supported by the Center for Lunar and Asteroid Surface Science (CLASS) under NASA cooperative agreement #80NSSC19M0214.